\documentstyle[twocolumn,aps,prl,epsf]{revtex}
\begin{document}
\draft
\date{\today}
\title{Non-hermiticity in a kicked model:
Decoherence and the semiclassical limit}
\author{Indubala I. Satija
\footnote{web address: http://physics.gmu.edu/~isatija}}
\address{Department of Physics, George Mason University,
 Fairfax, VA 22030}
\author{Arjendu K. Pattanayak
\footnote{web address: http://physics.carleton.edu/Faculty/Arjendu} }
\address{Department of Physics and Astronomy,
Carleton College, Northfield, MN 55057}
\maketitle
\begin{abstract}

We study the effects of non-hermitian perturbation on a quantum kicked
model exhibiting a localization transition. Using an exact
renormalization scheme, we show that the critical line separating the
extended and localized phases  approaches its semiclassical limit as
the imaginary part of the kicking parameter is steadily increased.
Further, the metastability of the quantum states appears to be
directly correlated with the deviation between the semi-classical and
quantum results. This direct evidence of quantum-classical 
correspondence
suggests that non-hermitian perturbations may be used to model 
decoherence.
\end{abstract}

%\narrowtext

Decoherence, namely the loss of quantum coherence through perturbations
from environmental degrees of freedom, is one of the most exciting and
active area of research at the forefront of physics. In addition to
its importance at technological fronts such as quantum
computing\cite{qcomp}, the subject is crucial to the field of quantum
chaos. It is argued that decoherence is essential in establishing the
correspondence principle for quantum systems with classically chaotic
limits\cite{zurek}. The subject has created theoretical and experimental
challenges in modeling and controlling the interaction of
quantum systems with the environment.

There are recent suggestions that decoherence and dissipation may be
modeled as non-hermitian perturbations\cite{ferry}. At the same time,
there has been a considerable amount of interest in delocalization
effects in non-hermitian systems, although these have not considered
the connection to decoherence at all. These studies were triggered by
the result\cite{nelson} that a complex vector potential delocalizes
wave functions otherwise localized in a random potential.
This is reminiscent of the delocalization due to decoherence of
dynamical localization in quantum systems with classically chaotic
dynamics\cite{raizen}. In this paper, we argue that effects of
non-hermiticity may be understood as a special case of the general
effect of quantum properties being destroyed by decoherence.
Among other interesting questions, this opens up the issue of
characterizing the transition from quantum to classical properties as a
function of the strength of the complex perturbation.

Earlier studies relating non-hermititian perturbations and
delocalization were confined to the complex vector potential. The
non-hermitian vector potential was argued to be intrinsically different
from non-hermitian scalar potentials\cite{brouwer} in that the
imaginary vector potential singles out a direction in space, breaking
the symmetry between the left and the right-moving particles, while the
imaginary scalar potential can be understood as singling out a
direction in time. However, our results indicate that transport
characteristics are {\em in general} affected by non-hermiticity
{\em irrespective} of where the non-hermitian terms appear. This line 
of
reasoning emerged from a recent study of non-hermitian lattice models
exhibiting a localization-delocalization transition\cite{JS}. There,
the non-hermitian scalar and vector potentials correspond to the
non-hermitian diagonal and off-diagonal perturbations which are related
by a Fourier transformation. That is, the effects on spatial 
localization
due to the complex vector perturbation correspond to the effects on
momentum space localization due to the scalar term. In particular, it
was found that for a complex vector potential, the extended phase is
accompanied by complex eigenenergies (as found earlier\cite{nelson})
while for the complex scalar potential the same was true for the 
localized
states. Thus, the main issue is the non-hermiticity itself rather than
its source in the vector or the scalar potential. This perspective
specifically interprets non-hermitian terms via their effects as
decohering perturbations. In the following, we study a non-autonomous
system, with non-hermitian kicking, exhibiting a 
localization-delocalization
transition. Using a renormalization group(RG) technique, we study the
effects of non-hermiticity on this transition. We show explicitly that
as the non-hermitian perturbation is increased, the system's
localization-delocalization phase diagram monotonically approaches
the semiclassically determined diagram.

Periodically kicked Hamiltonian systems such as
\begin{equation}
H=T(p) + V(x) \sum \delta(t-n)
\end{equation}
with $T(p)=p^2/2$ (kicked rotor) and with $T(p)=L \cos(p)$ (kicked
Harper) are an important class of theoretical and experimental
systems for studying the quantum dynamics of classically
non-integrable systems\cite{kmodel}. Despite extensive study for
almost two decades, questions of classical-quantum correspondence
and dynamical localization in these systems remain open.
When the quasienergy states of the system are projected on the angular
momentum basis, these models map onto lattice models\cite{Fishman}.
However, in contrast to the lattice models for autonomous systems,
kicked systems describe long-range interactions and hence are more
difficult to study. A special class of kicked models with
$V(x)= 2\hbar \arctan(\bar{K} cos(x))$ are useful\cite{Fishman,SS}
as they can be represented by a nearest-neighbor (NN) tight-binding
model (tbm) of the form
\begin{equation}
\psi_{m+1}+\psi_{m-1}+2/\bar{K} \tan[T(p)+\omega/2] \psi_m = 0
\end{equation}
where $\bar{K}=K/(2 \hbar)$ and $\omega$ is the quasi-energy.
Here the lattice index $m$ represents the angular momentum
quantum number in the absence of the kicking term.

With $T(p)=p^2/2$, this model corresponds to a lattice model with a
pseudo-random potential exhibiting localization with the localization
length equal to that of the Lloyd model\cite{Fishman,SS}. We study the
model with $T(p)=Lcos(p)$ which exhibits both extended and localized
phases\cite{SSunp} for irrational $\hbar/2\pi$. The system also 
describes
the NN truncation of the kicked Harper model and for small $\bar{K}$
and $\bar{L}$, it reduces to the Harper equation\cite{Harper} with
$E=\omega$. Finally, the system also describes the quantum
spin-$1/2$ XY chain, kicked periodically by a transverse magnetic
field\cite{tp}. This model is thus a good testing ground for 
investigating the effects of non-hermitian perturbations on the 
transport characteristics of non-autonomous systems.

Although this model has no nontrivial classical limit, there exists a
nontrivial semiclassical limit. As $\hbar\to 0$ the lattice
representation of the model can be written as the quantum continuum
Hamiltonian\cite{Wilkinson,SSunp,SSprl},
$H_{cont}= [\bar{K} \cos(p) + \tan[\bar{L} \cos(x+\omega/2)] =0$ where
$x=\hbar n$ and $[x,p] = i\hbar$. Remarkably, if this Hamiltonian is
interpreted classically, the resulting orbits carry a signature of
the localization-delocalization transition for the original lattice 
model.
Specifically, unbounded phase-space trajectories correspond to extended
states along $x$ or $p$ while a bounded orbit describes the
localization-delocalization boundary. It is easy to show that for this
continuum Hamitonian, such a bounded orbit exists for
$\bar{K}=\tan(\bar{L})$, independent of $\omega$. This is therefore the
semiclassical condition for the critical line separating the extended
and localized states. We show below that this semiclassical critical 
line is a reasonable approximation to the `exact' quantum critical 
line, and for complex kicking, the exact critical line tends toward 
this semiclassical line. For the Harper equation, which describes an 
autonomous system, the semiclassical prediction for the critical line 
is {\em exact}.

The quantum phase diagram in $K-L$ space is studied using a
renormalization group(RG) approach for a fixed quasi-energy. We use
dimer decimation\cite{GS,PSS} which has conceptual and intuitive
advantages over other RG schemes\cite{ostlund,ketoja}. The key idea
underlying the renormalization scheme is the {\em simultaneous}
decimation of the two central sites of the doubly infinite lattice
$-\infty, ... -2,-1,0,1,2,....\infty$, namely $\pm 1$,
$\pm 2$ and so on, after we have eliminated the central site $m=0$.
At the $n^{\rm th}$ step where all sites with $|m| < n$ have been
eliminated, the tbm for $m=\pm n$ reads
\begin{eqnarray}
\Phi_{n+1} + G^{+}(n) \Phi_{-n} - E^{-}(n) \Phi_{n}&=&0\\
\Phi_{-n-1} +G^{-}(n) \Phi_{n} - E^{+}(n) \Phi_{-n}&=&0
\label{RGv} \end{eqnarray}
where $G^x(n)$ and $E^x(n)$ ( $x=\pm$ ) respectively
describe the renormalized coupling and the on-site potential term
at the $n^{th}$ step of the renormalization. The $+(-)$ correspond
to  the right (left) parts of the lattice. With the initial conditions
determined by Eq.~(2), the renormalized parameters are given by the
{\em exact} RG flow\cite{GS}
\begin{equation}
{\bf M}_{n+1} = {\bf f}_{n+1} + {\bf M}^{-1}_{n}
\end{equation}
where ${\bf M}$ is a $2 \times 2$ matrix defined as
\begin{equation}
{\bf M} = \pmatrix{ E^{+} &  G^{-}\cr
             G^{+} &  E^{-}}
\end{equation}
and ${\bf f}$ is a diagonal $2 \times 2$ matrix, $f_{n,n}= E(n)$.
Asymptotically, the renormalized lattice can be viewed as a
dimer where the transport characteristics are determined by the quantum
interference between the two sites of the dimer. Interestingly, the
extended phase corresponds to a rigid dimer while the localized phase is
asymptotically a broken dimer. Therefore, the transport and localization
properties are described by the effective coupling of the renormalized
dimer, which is the ratio between the off-diagonal and the diagonal
part of the renormalized tbm $R(n) = G  G^{\dag}/ E E^{\dag}$. The
scaling exponent $\beta = \lim_{n\to\infty}\log R(n)/\log n$
effectively quantifies the transport properties, since extended states
are described by (typically monotonic) convergence of $\beta(n)$ $\to
0$. For exponential localization, $\beta(n) \to -\infty$. In
contrast, the critical states are characterized by negative $\beta$ and
non-convergent, oscillatory behavior. This provides an extremely high
precision method of obtaining a phase diagram for fixed quasienergy of
systems exhibiting extended, localized and critical states.

Figure~(1) shows the results, at almost machine precision, of this
method applied to our system\cite{fn}. The important feature to note
is that as the kicking parameter increases, the diagram has a narrow
re-entrant phase (a peak) and a plateau. Interestingly, with the 
exception of the region near this peak, the phase diagram is more or 
less consistent with the semiclassical prediction. Further, Fig.~(2) 
shows the transport characteristics for wavepackets in this system, and 
this global phase diagram corresponds closely to Fig.~(1), which is for 
a pure quantum state. This is important, since RG tools to compute the 
phase diagram for $\omega=0$ require a small fraction of the 
computational time for computing the wavepacket transport 
characteristics.

\begin{figure}
\begin{center}
\leavevmode
\epsfxsize=3in
\epsfbox{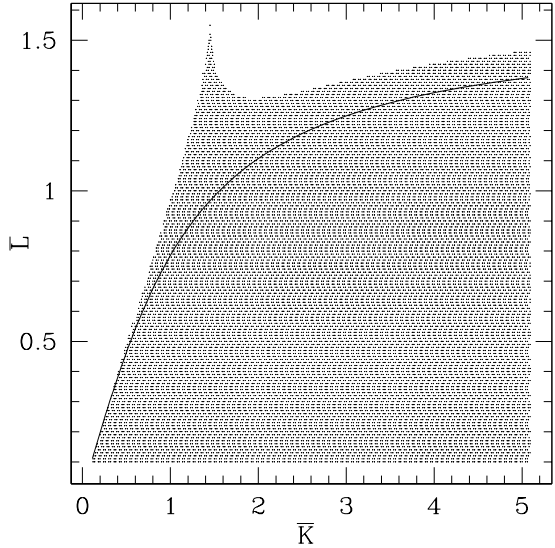}
\end{center}
\protect\caption{
Phase diagram for the kicked model with $\hbar/2\pi = (\sqrt(5)-1)/2$
and $\omega=0$. The shaded regime is the extended phase.
The solid line is the semiclassical critical line.}
\label{fig1}
\end{figure}

\begin{figure}
\begin{center}
\leavevmode
\epsfxsize=3in
\epsfbox{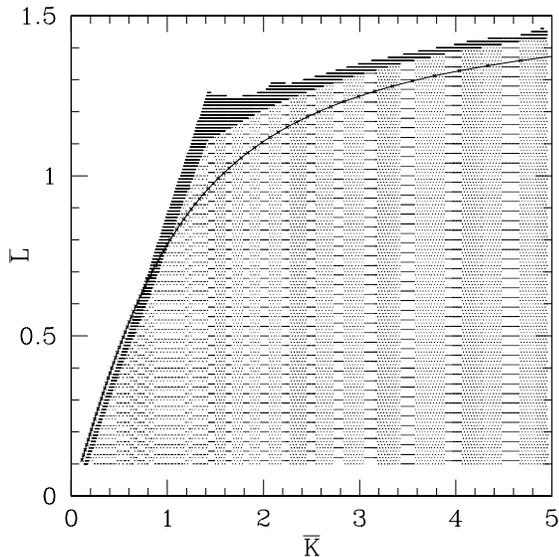}
\end{center}
\protect\caption{Transport properties of a quantum wave packet.
Using a plane wave initial condition, we compute $<p^2>$ after
1000 kicks. The lightly-shaded regime shows the parameter space
where $<p^2>$ is greater than $10^4$. The darker regime corresponds
to $<p^2>$ between $10^3- 10^4$. The narrowness of this regime
confirms that the computation is converging accurately to the
localization boundary.}
\label{fig2}
\end{figure}

In Fig.~(3) we show the effects of a complex perturbation,
$\bar{K} \to \bar{K}_r+i\bar{K}_i$, on the phase diagram. The
focus is on the changes in the critical line as $\bar{K_i}$ is 
increased.
What is interesting is the manner in which the transition curve
moves toward the semi classical critical line as the non-hermiticity
of the perturbation is increased. The transition appears to happen in
distinct stages: (a) the peak diminishes and then disappears,
(b) the curve gets closer in shape to the semiclassical line while still
maintaining a difference and finally(c) when the real part of
$\bar{K}$ is switched off, the curve is almost exactly on top
of the semiclassical line. Also, note that the effects of 
non-hermiticity are consistent with earlier work\cite{nelson}, since 
the non-hermitian perturbation shifts the critical line in parameter 
space so as to increase the measure of parameter space corresponding to 
dynamical localization (in momentum space) and hence enhance the 
parameter space corresponding to delocalization in real space. This 
reinforces our earlier arguments on delocalization being a phenomenon 
that occurs independent of the source of non-hermiticity.

\begin{figure}
\begin{center}
\leavevmode
\epsfxsize=3in
\epsfbox{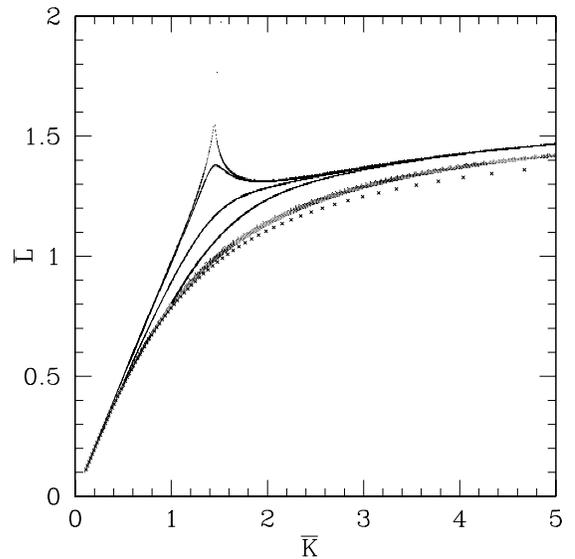}
\end{center}
\protect\caption{The effects of complex perturbation on the critical
line. The parameter $\bar{K}$ is the absolute value of the complex
kicking. The lines from top to bottom correspond to
$\bar{K}_i=0$,$\bar{K}_i=.1$, $\bar{K}_i=.5$, $\bar{K}_i=1$,
and finally with $\bar{K}_r=0$, respectively. The crosses show
the semiclassical critical line.}
\label{fig3}
\end{figure}

The RG approach was also used to study the effects of non-hermiticity on
the quasienergy spectrum\cite{PSS}. It turns out that with the exception
of the $\omega=0$ state, all other quasienergies are 
complex\cite{bender}.
The pure imaginary part of the spectrum exhibits a {\em band structure},
as shown in Fig.~(4). We show only {\em extended} states that are easy 
to compute from the RG analysis\cite{PSS}. The non-hermiticity thus 
associates continuum families of life-times with the state with 
$\omega=0$. As we approach the localization-delocalization boundary, 
this band splits into sub-bands with the notable localization of the 
$\omega_i=0$ band at the onset of localization as confirmed by further 
simulations. Therefore, the localization threshold is signalled by the 
$\omega_i=0$ band degenerating to a point and the state is localized, 
with both the real and imaginary part of the spectrum being point-like.

An intriguing feature of this pure imaginary part of the spectrum is the
presence of relatively large values of $\omega_i$ in the parameter
regime corresponding to the {\em peak} of the phase diagram Fig.~(1). 
The results shown in Fig.~(4) suggest that this regime, which 
corresponds to extended states, is more unstable than the rest of the 
delocalized phase. Since this part of the parameter space corresponds 
to greater deviation between the quantum and the semiclassical 
behavior, this relates the set of allowed life times of a quantum state 
to its proximity to the part of the phase diagram which is {\em not} 
described semi-classically. This is consistent with the intuition from 
the decoherence literature, and the statement that the part of the 
phase diagram which is not described by the semiclassical theory is 
most sensitive to the non-hermitian perturbation is arguably general.

\begin{figure}
\begin{center}
\leavevmode
\epsfxsize=3in
\epsfbox{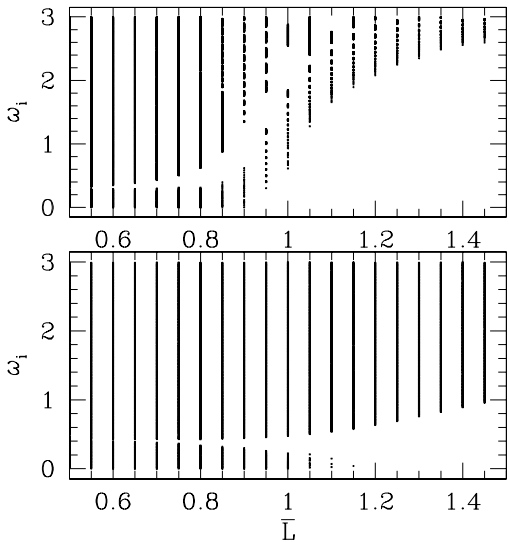}
\end{center}
\protect\caption{The pure imaginary part of the spectrum for 
$\bar{K_r}=0$ as
the localization transition is approached, for $\bar{K_i}=1.5$ (top) and
$\bar{K_i}=2.5$(bottom). The figure shows only extended states with
$|\omega_i| < 3$; the spectrum is symmetric about $\omega_i=0$.
A comparison between the top and the bottom figures shows that peak
part of the phase diagram is associated with shorter lifetimes and hence
is more unstable.}
\label{fig4}
\end{figure}

Previous work has indicated that complex Hamiltonians may be associated
with decoherence effects\cite{ferry,josh,audretsch}. The signature of
decoherence is that the quantum system behaves increasingly classically
as the decoherence parameter is increased, independent of the value
of $\hbar$. Decoherence is in fact argued to be necessary for
quantum-classical correspondence in chaotic systems\cite{zurek,ap}.
Our results are consistent with this, showing that results from
decoherence may be used to understand complex Hamiltonians, and
alternatively, that a complex Hamiltonian formulation may be used to
model decoherence effects. This opens new possibilities in modelling 
the
interaction between non-integrable systems and the environment,
providing a simpler alternative to solving master 
equations\cite{Salman}.
We hope that our study will stimulate further exploration of 
non-hermitian systems, and particularly of those displaying chaos.

We thank Tomaz Prosen and Nausheen Shah for their assistance in certain
aspects of the work. The research of IIS is supported by a grant from
National Science Foundation DMR~0072813.


\begin{references}

\bibitem{qcomp} See for example M. A. Nielsen and I.L. Chuang, {\em
Quantum Computation and Quantum information},(Cambridge University
Press, New York, 2000).

\bibitem{zurek} W. H. Zurek and J. P. Paz, Physica D, {\bf 83} 300
(1995).

\bibitem{ferry} D. Ferry and J. R. Barker, Appl. Phys. Lett., {\bf 74},
582 (1999).

\bibitem{nelson} N. Hatano and D. R Nelson, Phys. Rev. Lett.
{\bf 77}, 570 (1996); Phys. Rev. B {\bf 56}, 8651 (1997).

\bibitem{raizen} H. Ammann {\em et al}, Phys. Rev. Lett. 80, 4111
(1998);
B. G. Klappauf {\em et al} Phys. Rev. Lett. 81, 1203 (1998)

\bibitem{brouwer} P.W. Brouwer, P. G. Silvestrov and C. W. J. Beenakker,
Phys. Rev. B, {\bf 56} R4333 (1997).

\bibitem{JS} A. Jazeri and I. Satija, Phys. Rev. E,63,036222-1, 2001.

\bibitem{kmodel} {\it Quantum Chaos: between order and disorder: a
selection of papers}, edited by G. Casati and B. V. Chirikov
(Cambridge University Press, New York, 1995).

\bibitem{Fishman} See S. Fishman, D. R. Grempel, and R. E. Prange,
Phys. Rev. Lett. {\bf 49}, 509 (1982); Phys. Rev. A, {\bf 29}, 1639
(1984).

\bibitem{SS} I. Satija, B. Sundaram and J. Ketoja, Phys. Rev. E,
{\bf 60}, 453 (1999).

\bibitem{Harper} P. G. Harper, Proc. Phys. Soc. London A {\bf 68}, 874 
(1955).

\bibitem{tp} T. Prosen, Phys. Rev. E, {\bf 60}  1658 (1999).

\bibitem{Wilkinson} M. Wilkinson, Proc. R. Soc. London A391, 305 (1984).

\bibitem{SSunp} For hermitian kicking, this model was investigated
by I.~Satija  and B.~Sundaram (1998,unpublished).

\bibitem{SSprl} I. Satija and B. Sundaram, Phys.~Rev.~Lett.{\bf 84},
4581 (2000).

\bibitem{GS} I.~Gomez and I.~Satija, Phys.~ Lett. A, {\bf 268} 128 
(2000).

\bibitem{PSS} T.~Prosen, I.~Satija and N.~Shah, Phys.~Rev.~Lett.
{\bf 87}, 066601 (2001); I. Satija and T Prosen, 2002 (preprint).

\bibitem{ostlund} S. Ostlund and R. Pandit, Phys. Rev. B {\bf 29}, 1394
(1984);

\bibitem{ketoja} J. A. Ketoja, Phys. Rev. Lett. {\bf 69}, 2180 (1992).

\bibitem{fn} Our studies concentrate on the region $\bar{L}<\pi/2$
where the semiclassical predictions are expected to be valid.

\bibitem{bender} The fact that $\omega=0$ continues to exist in real
space may be a consequence of some special symmetry of this state.
See for example C. Bender and S. Boettcher, Phys. Rev. Lett., {\bf 80}
5243 (1998).

\bibitem{Harper} P. G. Harper, Proc. Phys. Soc. London A {\bf 68}, 874 
(1955).

\bibitem{josh} J. Wilkie, J.Chem. Phys. {\bf 115}, 10335 (2001).

\bibitem{audretsch} J. Audretsch and M. Mensky, \pra {\bf 56}, 44
(1997).

\bibitem{ap} A.K. Pattanayak, Phys. Rev. Lett. {\bf 83}, 4526 (1999).

\bibitem{Salman} S. Habib et al., Phys. Rev. Lett. 80, 4361 (1998).

\end{references}
\end{document}